\documentclass{svmult}
\usepackage{type1cm,graphicx,newtxtext,newtxmath,csquotes,xcolor,enumitem}
\usepackage[bottom]{footmisc}
\usepackage[style=biblatex-spbasic,natbib=true,uniquename=init,dashed=false,maxbibnames=6,doi=false]{biblatex}
\bibliography{Arroyo.bib}

\usepackage[colorlinks,allcolors=blue]{hyperref}
\hypersetup{
  pdftitle={Non-individuality and experience},
  pdfauthor={Raoni Arroyo}
}

\title*{\texorpdfstring{Non-individuality and experience\thanks{Invited contribution to appear in \fullcite{arenhartkrause2024}. Manuscript version: \today.}}{Non-individuality and experience}}
\author{Raoni Arroyo}
\institute{Raoni Arroyo \at Centre for Logic, Epistemology and the History of Science (CLE) \at Graduate Program in Philosophy, University of Campinas (UNICAMP), Campinas, Brazil. \at Research Group in Logic and Foundations of Science (CNPq), Florianópolis, Brazil \at \email{raoniarroyo@gmail.com}}

\motto{---And the funny thing is, our cells are completely regenerating every seven years. We've already become completely different people several times over, and yet we always remain quintessentially ourselves.\\(Waking Life, 2001)}

\begin{document}

\maketitle

\abstract{This chapter acknowledges a gap between the ``non-individuals'' interpretation of quantum mechanics and our world of experience, and begins to bridge it. \S \ref{sec:1} states the problem with Abner Shimony's ``Phenomenological principle''; \S \ref{sec:2} briefly presents the interpretation with connection to standard quantum mechanics; \S \ref{sec:3} presents the measurement problem in connection with the Phenomenological principle, the standard way out of it, and why the ``non-individuals'' interpretation of quantum mechanics should not follow it; \S \ref{sec:4} finally shows two closed venues for such an interpretation (Bohmian mechanics and the Modal-Hamiltonian Interpretation), and two alternatives for such it (Everettian quantum mechanics and spontaneous collapse theories).}

\keywords{
Non-individuals;
Non-relativistic quantum mechanics;
Phenomenological principle;
Quasi-set theory.
}

\section{The phenomenological world of experience}\label{sec:1}

The epigraph is an excerpt from Richard Linklater's 2001 movie, ``Waking Life'', in a scene known as the ``Coffee Shop'' or the ``aging paradox'' scene. While the whole scene is one of my favorites, the line quoted above suffices as a starting point for our discussion: we are used to going through life and thinking about individuals. Yet, this is precisely what the ``non-individuals'' interpretation of quantum mechanics  (QM$^{\text{NI}}$ hereafter) disrupts.\footnote{What we are calling QM$^{\text{NI}}$ is also known as the ``Received View'' \citep{arenhart2017} of quantum individuality.}

Our entire legal system is also based on that. Just imagine the chaotic situation in which one cannot be legally responsible for their actions because one is not the same person who committed such an action. All our personal life is also based on the idea that we know and love specific individuals. Different examples are all over the place. Were Neymar to hit Brazil's penalty kick in the World Cup 2022, maybe that match's score would have been different. And so forth and so on. Everything that happens in our world mostly depends on specific individual persons performing specific tasks; were those persons replaced by different persons, the outcome of the task would have the possibility of yielding a new physical situation.

In Western philosophy, it all traces back to the problem of the Ship of Theseus: after coming and going, eventually all parts of the ship are replaced, so how come that's the same ship?  However, I bet Theseus would have been concerned if the ship got robbed, with or without compelling reasons of the vessel being his own. Because that's how the world of \textit{experience} runs.

QM$^{\text{NI}}$ gives a completely different picture of how the world could be. According to it, quantum entities, interpreted as \textit{objects}, fail to have an individuality profile. Most recently, the QM$^{\text{NI}}$ understood the notion of ``object'' as a ``nomological object'', which is a sense first introduced by \citet{toraldo1978} in this context \citep[see][]{arenhart2023,krausejorge2024}. It forces our understanding of the world from the ground up: the world is \textit{made up} of non-individual objects. Fair enough. But, in doing so, it lacks what \citet[147]{shimony1997} called the ``Phenomenological principle'':
\begin{quote}
    \textelp{} whatever ontology a coherent philosophy recognizes, that ontology must suffice to account for appearances. \citep[147]{shimony1997}
\end{quote}
That is, it lacks a story of how non-individual entities could make up individual entities. While \citet[147]{shimony1997} takes the principle to be sufficient to falsify metaphysical theories that don't cope with it (his particular example is eliminativist metaphysics with respect to the human mind), I wouldn't go this far by saying that QM$^{\text{NI}}$ should be ruled out because it lacks such a principle. The challenge is, then, to offer a Phenomenological principle to QM$^{\text{NI}}$ so it could tell us a story about how our world of appearances/experience can be derived from the story of atoms and electrons.

In the foregoing, we'll see how such an interpretation might answer to this in order for it to advance metaphysically.

Of course, there are two sets of problems at stake here. Firstly, there is the problem of diachronic individuality, or \textit{individuality over time}\footnote{Sometimes written \textit{identity} over time, a terminology I want to avoid in this chapter to prevent misunderstandings between the logical notion of identity and the metaphysical notion of individuality; for details on this distinction, see \citet{krausearenhart2018}.} \citep{sep-identity-time}. Then there is the problem of synchronic individuality, which is the problem of accounting for individuality at any given instant of time. Naturally, both diachronic and synchronic individualities are a problem for quantum entities; but as the former is a problem that applies to several objects in our world (\textit{e.g.}, ships, persons, clouds, \textit{etc.}), the latter is an exclusivity of quantum entities.

\section{The ``non-individuals'' interpretation}\label{sec:2}

QM$^{\text{NI}}$ is often presented not as an interpretation of quantum mechanics in the sense of a solution to the measurement problem, but as a particular metaphysical reading of the standard interpretation of quantum mechanics (or ``standard quantum mechanics'').
\citet{krausearenhartbueno2022} summarize the methodology of QM$^{\text{NI}}$ as follows:

\begin{quote}
    If [standard] quantum mechanics is understood as dealing with objects of a given kind, whether particles, fields, or something else, it may be asked: what are these objects metaphysically? This, in turn, leads to questions regarding whether they are individuals or not, and if they are, which principle of individuality determines that that is the case? The non-individuals interpretation of quantum mechanics [QM$^{\text{NI}}$] takes the relevant entities as lacking individuality, adding a further metaphysical interpretative layer over the theory's bare entities. \citep[1136]{krausearenhartbueno2022}
    \end{quote}
    
The traditional case in point is the Bose--Einstein statistics (Table \ref{tab:statistics}), in which one has three possible ways of organizing two particles in two boxes. The permutation of particles in case 3 below doesn't yield a new physical situation \textit{because} they're objects lacking individuality.

\begin{table}[ht!]
\centering
\caption{Statistics for particles in boxes}
\label{tab:statistics}
\begin{tabular}{lcc}
\hline\noalign{\smallskip}
& \textbf{Box $1$} & \textbf{Box $2$}\\
\noalign{\smallskip}\hline\noalign{\smallskip}
1.&$\bullet\bullet$ &     \\
2.& & $\bullet\bullet$    \\
3.&$\bullet$ & $\bullet$  \\
\noalign{\smallskip}\hline
\end{tabular}
\end{table}

The relevant thing for QM$^{\text{NI}}$ is that it collapses the lack of \textit{individuality} and lack of \textit{identity} by virtue of the way in which the individuality profile is understood. For instance, haecceity is a non-qualitative property that individualizes someone/something. This is also called ``transcendental individuality'' due to \citet{post1963}. The haecceistic property of someone/something is ``to be identical with someone/something''. In this way, the haecceistic property of my laptop is ``being \textit{identical} with itself''; we might find countless indiscernible laptops with the same brand, color, configuration, \textit{etc.}, but none of them---except mine---has the non-qualitative property of being identical with this particular laptop, \textit{i.e.}, \textit{itself}. And the same applies to all other laptops; each one of them has the property of being \textit{identical with itself}, and this individualizes each laptop. The same would be true for Ship of Theseus: it has the property ``$\{\text{Ship of Theseus}=\text{Ship of Theseus}\}$'', and this non-qualitative, haecceistic property remains with it even though all parts of the ship are eventually replaced. Hence, this is how individuality and identity collapse in QM$^{\text{NI}}$:

\begin{quote}
\textelp{} the idea is apparently simple: regarded in haecceistic terms, ``Transcendental Individuality'' can be understood as the identity of an object with itself; that is, ``$a=a$''. We shall then defend the claim that the notion of non-individuality can be captured in the quantum context by formal systems in which self-identity is not always well-defined, so that the reflexive law of identity, namely, $\forall x(x=x)$, is not valid in general. \citep[13--14]{frenchkrause2006}
\end{quote}

Methodologically, this is the importance of a quasi-set theory, that is, a theory which enables one to talk about the lack of individuality right from the start. As \citet{frenchbigaj2024} nicely summarize, there are two ``basic posits'' (``Urelemente'') in quasi-set theory:
\begin{itemize}
    \item $M$-atoms. They compose all everyday objects such as tables, chairs, persons, and laptops. To them, the concept of identity ($=$) and---crucially---the concept of \textit{individuality} apply. It is just like classical set theory with Ur-elements. In this sense, they're ``classical'' objects, both in logical and physical sense. Objects of this kind, such as you and I, are $M$-objects.
    \item $m$-atoms. They compose ``quantum'' objects such as electrons and fields. For them, the usual identity and individuality concepts don't apply, but mere indistinguishability ($\equiv$) which is a weaker notion. It is because of $m$-atoms that quasi-set theory is a non-classical logical (\textit{qua} non-reflexive) system, as the identity principle fails to apply universally. Objects of this kind, such as electrons, are $m$-objects.
\end{itemize}

Crucially, for the notion of non-individuality, quasi-set theory, as \citet{frenchbigaj2024} put it
\begin{quote}
    \textelp{} supply the beginnings of a categorial framework for quantum ``non- individuality'' which, it is claimed, helps to articulate this notion and, bluntly, make it philosophically respectable
\end{quote}

So the world is divided into two kinds of entities, the individual ones and the non-individual ones. And---because of the way in which metaphysical individuality is cashed out---individuality is tied up with identity.

This immediately raises the question: when does a $m$-object become a $M$-object? If quantum objects make everyday objects, there should be a threshold after all. An answer to that would be an answer to the Phenomenological principle, and this is of the utmost importance for a metaphysical view. The answers, however, aren't clear. Such a Phenomenological principle is captured by the ``Crisp axiom'' (C) of quasi-set theory---where C is a predicate that turns the $m$-atom $x$ into a $M$-atom $x$, \textit{e.g.}, a classical entity---in which:

\begin{equation}\tag{C}
    \forall x \bigg(\Big(m(x)\to\big(C(x)\to M(x)\big)\Big)\bigg)
\end{equation}

\citet{krause2012} has developed this idea in an unpublished manuscript that only recently caught philosophers' attention with regards to this specific axiom \citep[see][p.~122 ff.]{maciasbustos-martinezordaz2023}:
\begin{quote}
    A certain $m$-object may, by some process, described case-by-case by a device described by the predicate $C$, becomes a $M$-object, say when a quantum entity becomes ``classical'', making a click in an experimental device. Thus, $C$ stands for ``crisp'', in opposition to ``blurring''. Intuitively speaking, $C$ so to say eliminates the quantum behaviour of $x$. It expresses a kind of ``collapse'' of something related to the quantum entity. \citep[9]{krause2012}
\end{quote}

While something such as axiom C is crucial to fill the gap required by the Phenomenological principle, something looks like out of place. In particular, the crucial question that remains unanswered is that, even if we fix the Phenomenological principle, how come an individual (a $M$-atom) can be formed by a collection of non-individuals ($m$-atoms)? We have seen this kind of divide before in quantum mechanics, and it let to an instance of the (in)famous \textit{measurement problem} of quantum mechanics. So things do not look good from where they stand now. Let us see why.

\section{The measurement problem's problem}\label{sec:3}

It is widely known that the measurement problem has haunted quantum mechanics since its early foundational debates. It is also widely known that it can be formulated in several ways, from dilemmas \citep{bell1989} to trilemmas \citep{maudlin1995} to polylemmas \citep{muller2023}. The simplest case is straightforward: Schrödinger's cat enters a box, and it turns out that after one hour the cat might be in state $|A\rangle$ or in state $|B\rangle$, and quantum mechanics says it is in state $a|A\rangle+b|B\rangle$---which is a vector sum that yields, according to the Born Rule, $|A\rangle$ as a measurement outcome with probability $|a|^2$ and $|B\rangle$ with probability $|b|^2$---that isn't translatable neither into
\begin{enumerate}
    \item $|A\rangle$
    \item $|B\rangle$
    \item $|A\rangle\land|B\rangle$
    \item $\neg|A\rangle\land\neg|B\rangle$
\end{enumerate}
but a state of its own.

Now there's a more basic question: why must we fix this? Namely, why is the measurement problem a problem in the first place? And the answer is Shimony's so-called ``Phenomenological principle''. The measurement problem is a problem \textit{because} we don't experience superposition of macroscopically distinguishable states such as cat states $|A\rangle$ and $|B\rangle$. It messes up with how we perceive our world of experience \citep{albert1992}. That's why we need to fix this.

One very popular way to fix this situation is via Standard Quantum Mechanics (SQM), or the so-called ``Copenhagen'' interpretation. This strategy consists of positing a \textit{deus ex machina} process---the \textit{collapse}---that makes the transition $a|A\rangle+b|B\rangle$ to either $|A\rangle$ or $|B\rangle$ with the aforementioned probability measure due to some sort of interaction---the \textit{measurement act} \citep{vonneumann1932}---between the quantum system and another non-quantum system such as a measurement apparatus \citep{bohr1958} or a human consciousness \citep{wigner1961}. The point of this solution is to build a wall, so to speak, between the quantum and the non-quantum, such as the classical/quantum divide, or the macroscopical/microscopical, conscious/non-conscious, \textit{etc.}

The \textit{problem} with this kind of solution is that it doesn't really help us get one's head around this kind of problem as it bears on the---undefined---notion of ``measurement''. When a ``measurement'' takes place, a quantum system ceases to be described as a superposition of macroscopically distinguishable states and starts being described as a single, definite macroscopic state (such as $|A\rangle$ or $|B\rangle$). But what is a ``measurement''? That's what SQM doesn't tell. And that's why the concept seems to be a placeholder for whatever it is that makes the deus ex machina process of collapsing macroscopic superpositions into single/defined macroscopic states take place.

Of course, if one is an instrumentalist and only cares about measurement outcomes, then SQM is completely fine. It allows one to write quantum mechanics textbooks, develop technologies, and all sorts of applications we know and love. What it doesn't allow, however, is the possibility of one's deriving a picture of how the world could be according to SQM, as it deliberately moves away from these kinds of questions.

So \textit{if} this is one's goal, then SQM is not one's road to interpreting quantum mechanics---and, as I take it, QM$^{\text{NI}}$ has such a goal. However, as it seems, QM$^{\text{NI}}$ stand or fall with the SQM collapse postulate:

\begin{quote}
    It [quasi-set theory] takes standard quantum mechanics (and its extension to quantum field theory) as a fundamental physical theory. This is captured in their system by the introduction of the Crisp axiom. \textelp{U}pon measurement objects become Crisp. \citep[122]{maciasbustos-martinezordaz2023}
\end{quote}

But this cannot be the whole story if QM$^{\text{NI}}$ wants to attain its goal of describing a picture of how the world is (or how it \textit{could be}). This is why---or so I argue---QM$^{\text{NI}}$ needs to be attached with other formulations of quantum mechanics in order to make such a goal viable.

\section{\texorpdfstring{Venues for QM$^{\text{NI}}$}{Venues for QM-NI}}\label{sec:4}

A good place to start would be in the solutions to the measurement problem that as \citet[131]{daumer-etal2006} put it, ``\textelp{} do not postulate some special physics for measurements''. We'll focus on four research programs: the Modal-Hamiltonian Interpretation, Bohmian mechanics, Everettian quantum mechanics, and spontaneous collapse quantum mechanics. Let us briefly analyze them, in turn, to see how QM$^{\text{NI}}$ could (or couldn't) fit them.

The Modal-Hamiltonian Interpretation of quantum mechanics describes the domain of possibilities rather than relying on wavefunction collapse. In this framework, the Hamiltonian plays a central role, being used in the definition of systems and subsystems as well as in identifying the definite-valued observables \citep[1247]{dacostalombardi2014}. The quantum state, in this context, is interpreted as representing ontological propensities for the actualization of possible properties \citep[1248]{dacostalombardi2014}. Measurement is understood simply as an interaction that, as any interaction, leads to the actualization of a possible property into an actual property, according to the ontological propensities of the system given by the probabilities encoded by the Born Rule \citep[\S 2]{holik-etal2022}.

This interpretation is particularly compatible with the metaphysics of QM$^{\text{NI}}$, as it rejects the traditional conception of particles as individual entities endowed with individuality. Such an interpretation was, indeed, recently built upon the formalism of quasi-set theory \citep{holik-etal2022}. Instead, the Modal-Hamiltonian Interpretation adopts a property ontology, in which quantum systems are described as bundles of instances of universal type-properties, without any principle of individuality \citep[1249]{dacostalombardi2014}. These bundles consist of instances of universal type-properties (determinables) and their possible case-properties (determinates). This approach fully dispenses with the need for a substratum or a principle of individuality, positioning quantum systems as holistic collections of properties \citep[61--62]{lombardi2023}. Most recently, however, developments on the MHI have done away with the broader ontological category of ``object'' itself \citep[64--65]{lombardi2023}, thus creating a tension with QM$^{\text{NI}}$ as it presently stands---for instance, as presented in \citet{krausearenhartbueno2022}, which is built upon the ontological category of ``objects''. Hence, some modifications in the ontological categories of QM$^{\text{NI}}$ are needed to make it fully compatible with the most recent version of MHI.

On to the next one. In Bohmian mechanics, the solution to the measurement problem is to state that quantum mechanics is incomplete, as the quantum objects always have a pre-measurement definite position. Still, the quantum-mechanical description fails to describe what that definite position is. There's a perennial wave, being everywhere at the same time---called the ``pilot wave''---that drives particles instantaneously from place to place. Hence, the particle always has a definite position and trajectory. We don't know what its position/trajectory is due to our ignorance of the initial conditions from which it started out---but the important thing to be noticed is that it \textit{has} a definite value of position at all times. In this sense, a state such as $a|A\rangle+b|B\rangle$ depicts our ignorance, as the state is either $|A\rangle$ or $|B\rangle$. This way, the particle's individuality \textit{qua} spatio-temporal position is always conferred \citep{brown-etal1994,redhead1983,frenchbigaj2024}. Hence, not a venue for QM$^{\text{NI}}$.

But something interesting happens to this debate if Bohmian mechanics turn out to be the true description of nature. Were Bohmian mechanics true, then all quantum entities would have a definite position at all given moments. As mentioned earlier, this is one way of asserting an individuality profile to quantum entities. Hence, in the case of Bohmian mechanics turns out to be a true description of nature, then QM$^{\text{NI}}$ turns out to be a \textit{false} description of nature.\footnote{And this would be a case of the meta-Popperian methodology at work, with science constraining the metaphysical possibilities---see \citet{arenhartarroyo2021metaphilosophy} for details of such a methodology in metametaphysics.}

Let's try the next one. In Everettian quantum mechanics, the state  $a|A\rangle+b|B\rangle$ is read ontologically, which is to say, it doesn't represent our epistemic ignorance but how things are. So such a superposition of macroscopically distinguishable states would depict different branches of the universe; two separate ``worlds'' as it were. But the splitting into worlds is not an abrupt process just as SQM collapse, it doesn't happen in a definite moment of time. As \citet{durrlazarovici2020} have it,

\begin{quote}
    When we say that a world ``splits'' or ``branches'' (for instance, in the course of a measurement experiment), we are actually talking about a gradual process. Think of a wave packet on an extremely high-dimensional configuration space fanning out into two or more parts that become more and more separated in that space. Don't try to think of an exact moment in which it goes ``bing'' and the world suddenly multiplies. The concept of a ``world'' has a certain vagueness---it's not possible, in general, to say exactly how many worlds exist or at what moment in time a new splitting has occurred. \citep[p.~118]{durrlazarovici2020}.
    \end{quote}
The probability measure represents then in which world we find ourselves within the multiverse: within the world in which the state of affairs $|A\rangle$ happens or $|B\rangle$. 

Individuality-wise, the non-individuality of $m$-atoms would, as soon as the world branches, become $M$-atoms, so this could be a nice venue for QM$^{\text{NI}}$. It just needs to modify C, from measurements to branching. This would be a more rigorous way of stating C, not being related to the SQM \textit{deus ex machina} collapse, as Everettian quantum mechanics has a more thorough way of saying how the world splits via decoherence \citep{wallace2012}. Crucially, however, the Everettian quantum mechanics solution must account for the Phenomenological principle by explaining why we don't see other worlds. The answer is that they're parallel, hence our $M$-atoms individualities are given by looking back at the ``tree'' of branching systems, so to speak.

Another venue for QM$^{\text{NI}}$ would be that of spontaneous collapse theories. It is known that due to \citet*{grw}, spontaneous collapse quantum mechanics introduces new constants of nature. A crucial one for our present purposes is the \textit{collapse rate} $\tau$. If $\tau$ is high enough, systems such as $a|A\rangle+b|B\rangle$ undergo a collapse to one of its components with a probability given by the probability measure $|a|^2$ or $|b|^2$. Moreover, $\tau$ increases with the degrees of freedom of the system under consideration, which is to say that the more complex the system is, the more likely to collapse to one of its macroscopically distinguishable states it is---which is what we expect to see in our world of experience. Hence, there is no problem with the Phenomenological principle here.

\section{Final remarks}

It isn't an easy job to tailor new metaphysical devices. Here's \citet{French2018} about such a task:
\begin{quote}
    What would be involved in constructing such ``new'' metaphysical devices? There is an obvious and immediate issue of language---I can introduce whatever new terms I like but if they're not relatable to familiar ones, I'm likely to receive a stare of incomprehension if not incredulity! And if they are so relatable, then any comprehension that follows will of course derive from that attached to whatever current metaphysical device the familiar term designates. \citep[227]{French2018}
\end{quote}
I take that French's claim can be read as an instance of Shimony's Phenomenological principle. We need the individuality that we experience---that's our \textit{familiar} device---to make sense of the metaphysical notion of non-individuality. And that's precisely why the ``non-individuals'' interpretation of quantum mechanics (QM$^{\text{NI}}$) must have the Crisp axiom (C), or some analog of it. C is needed to fill the Phenomenological principle's gap on the pains of---among other things---being incomprehensible to the human mind.

The introduction of C, however, won't do the entire job. As it was originally conceived, C played the role that ``measurement'' plays in Standard Quantum Mechanics (SQM). While the collapse makes a quantum system classical in SQM, C turns non-individual objects into individual objects in QM$^{\text{NI}}$. That's not what we wanted, however, as SQM is notoriously silent on what a measurement is, how and when it happens, and so on and so forth.

Simply put, one must have a narrative that explains the events or features that make the Crisp axiom applicable. This narrative should include aspects of the world that, on the one hand, are not fully determined by physics alone and, on the other, describe how the world could be, assuming quantum mechanics is true. That's why I have argued that we should detach QM$^{\text{NI}}$ from SQM, and attach it in one of the so-called interpretations of quantum mechanics. In a non-exhaustive sampling, I've listed four of them.

As QM$^{\text{NI}}$ currently stands, it is incompatible with Bohmian Mechanics and the Modal-Hamiltonian Interpretation (MHI). QM$^{\text{NI}}$ is irreconcilable with Bohmian mechanics due to individuality \textit{qua} spatio-temporal position, meaning that eventual future empirical testings favoring Bohmian mechanics would entail the falsification of QM$^{\text{NI}}$. With regards to MHI, the situation is quite different, as a bridge might be built. If future developments of QM$^{\text{NI}}$ adopt eliminativism regarding the ontological category of objects, then the compatibility with MHI is back on the table. The job is fairly easy when QM$^{\text{NI}}$ is attached to either Everettian quantum mechanics or spontaneous collapse theories. To further articulate the view with such interpretations is a task left for future metaphysics of science. 

QM$^{\text{NI}}$ thus stands at a crossroads. Once it sheds the---using the words of \citet{Einstein1928}---``tranquilizing philosophy'' offered by SQM, QM$^{\text{NI}}$ enters the \textit{interpretation wars} in quantum foundations. An immediate problem of that step away from tranquility is that justifying one's position in such a dispute may require---perhaps---another century of debate.

\section*{Acknowledgements}
I would like to thank Jonas Arenhart and Olimpia Lombardi for their comments on an earlier draft. This chapter was written during my research visit to the Department of Philosophy, Communication, and Performing Arts at Roma Tre University in Rome, Italy. It was supported by grants \#2021/11381-1 and \#2022/15992-8 from the São Paulo Research Foundation (FAPESP).

\printbibliography
\end{document}